\documentclass[11pt]{article}
\usepackage{amssymb}
\usepackage{amsmath}
\begin{document}
\abovedisplayshortskip 12pt
\belowdisplayshortskip 12pt
\abovedisplayskip 12pt
\belowdisplayskip 12pt
\baselineskip=15pt
\title{{\bf Hamiltonian Integrability of  Two-Component Short Pulse Equations}}
\author{J. C. Brunelli\thanks{\texttt{jcbrunelli@gmail.com}}  \\
\\
Departamento de F\'\i sica, CFM\\
Universidade Federal de Santa Catarina\\
Campus Universit\'{a}rio, Trindade, C.P. 476\\
CEP 88040-900\\
Florian\'{o}polis, SC, Brazil\\
\\
\\
S. Sakovich\thanks{\texttt{saks@tut.by}} \\
\\
Institute of Physics\\
National Academy of Sciences\\
220072 Minsk, Belarus\\
}
\date{}
\maketitle

\begin{center}
{ \bf Abstract}
\end{center}

We obtain the bi-Hamiltonian structure for some of the two-component short pulse equations proposed in the literature to generalize the original short pulse equation when polarized pulses propagate in  anisotropic media.
\bigskip



\newpage

\section{Introduction}

The nonlinear Schr\"{o}dinger (NLS) equation is one of the universal nonlinear integrable equation since it describes the slow modulation of the amplitude of a weakly nonlinear wave packet in a moving medium \cite{Benney1967}. It has been used with great success in nonlinear optics to describe  the propagation of sufficiently broad pulses, or slowly varying wave trains whose spectra are narrowly localized around the carrier frequency \cite{Hasegawa2010}. Today high-speed fiber-optic communication demands ultra-short pulses and technological progress for creating them was achieved, however,  in these conditions the description of the evolution of these pulses lies beyond the usual approximations leading to the NLS equation.  In \cite{Schafer2004} Sch{\"{a}}fer and Wayne proposed an alternative model to approximate the evolution of ultra-short intense infrared pulses in silica optics. After a scale transformation of variables their short pulse (SP) equation can be written as
\begin{equation}
u_{xt} =u +{1\over 6}\left(u^3\right)_{xx}\quad{\rm or}\quad u_{t} =\left(\partial^{-1}u\right) +{1\over 2}u^2u_x\;,\label{SP}
\end{equation}
with $u=u(x,t)$ representing the magnitude of the electric field and subscripts $x$ and $t$ standing for partial differentiations. It was shown in \cite{Chung2005} by numerical simulations that the SP equation is a success to describe pulses with broad spectrum.

In the last years the SP equation became a comprehensively studied equation of soliton theory. The integrability of this equation was first established in \cite{Sakovich2005} where for the linear spectral problem of the Wadati--Konno--Ichikawa type
\begin{equation}
\Phi_x=X\Phi\;,\quad \Phi_t=T\Phi\,,\label{spsp}
\end{equation}
the corresponding zero-curvature representation (ZCR)
\begin{equation}
X_t-T_x+\left[X,T\right]=0\;,\label{zc}
\end{equation}
is given by
\begin{equation}
X=\left(
                   \begin{array}{cc}
                     \lambda & \lambda u_x\\\noalign{\vskip 10pt}
                     \lambda u_x & -\lambda\\
                   \end{array}
                 \right)\;,\qquad
                 T=\left(
                   \begin{array}{cc}
                     {\lambda\over2}u^2+{1\over 4\lambda} & {\lambda\over6}\left(u^3\right)_{x}- {1\over2}u\\\noalign{\vskip 10pt}
                      {\lambda\over6}\left(u^3\right)_{x}+ {1\over2}u& -{\lambda\over2}u^2-{1\over 4\lambda} \\
                   \end{array}
                 \right)\;.\label{zcsp}
\end{equation}
Also in \cite{Sakovich2005} a transformation was found which relates the SP to the sine-Gordon equation and it was used in \cite{Sakovich2006} to derive exact soliton solutions of the SP equation from known soliton solutions of the sine-Gordon equation. In \cite{Sakovich2007} the SP equation was studied as one of four Rabelo equations of differential geometry  \cite{Rabello1989}. The recursion operator, Hamiltonian structures and conservation laws for the SP equation were also found in \cite{Brunelli2005,Brunelli2006}.The first Hamiltonian structure  follows from the Lagrangian density for the SP equation in nonlocal form (\ref{SP})
\[
{\cal L}={1\over 2}u\left(\partial^{-1}u_t\right) - {1\over24}u^4 + {1\over 2}\left(\partial^{-1}u\right)^2\;,
\]
and reads
\begin{equation}
u_t={\cal D}_1{\delta H_2\over\delta u}\quad{\rm with}\quad{\cal D}_1=\partial\;,\quad
H_2 = \int dx\left[{1\over24}u^4-{1\over2}(\partial^{-1}u)^2\right]\;.\label{fhssp}
\end{equation}
The second Hamiltonian structure follows, for instance, from the recursion operator obtained in \cite{Sakovich2005} using the cyclic basis technique \cite{Sakovich1995,Sakovich2004} and yields
\begin{equation}
u_t={\cal D}_2{\delta H_1\over\delta u}\quad{\rm with}\quad{\cal D}_2=\partial^{-1}+u_x\partial^{-1}u_x\;,\quad
H_1 = {1\over2}\int dx \,u^2\;.\label{shssp}
\end{equation}
Many other results were found concerning the SP equation such as Hirota's bilinear representation \cite{Kuetche2007}, multisoliton solutions \cite{Matsuno2007} and periodic solutions \cite{Parkes2008,Matsuno2008}, among others.

To take into account effects of polarization and anisotropy (see \cite{Kartashov2003} and references therein) two-component $(u,v)$ integrable generalizations of the SP equation were proposed in the literature \cite{Pietrzyk2008}-\cite{YaoZeng2011}. These equations play the same role in dynamics of ultra-short pulses as the pair of coupled NLS equation does for broader pulses \cite{Manakov1974}. Note that some of these systems reduce to the SP equation (\ref{SP}) if $u=0$ or $v=0$ while other systems if $u=v$. The integrability of these systems were obtained mainly from a zero curvature representation or bilinear formalism. An algebraic approach to the integrability based on a Hamiltonian formulation for infinite-dimensional dynamical systems \cite{Blaszak1998} is lacking  and in what follows we address the bi-Hamiltonian integrability of some of these systems written in evolutionary form
\[
\left(
                   \begin{array}{c}
                    \!\! u \!\!\\\noalign{\vskip 5pt}
                     \!\!v \!\!\\
                   \end{array}
                 \right)_t={\cal D}_1
\left(\begin{array}{c}
                    \!\!  {\displaystyle{\delta H_2/\delta u}}\!\!  \\\noalign{\vskip 5pt}
                     \!\! {\displaystyle{\delta H_2/\delta v}}\!\!  \\
                   \end{array}\right)={\cal D}_2
\left(\begin{array}{c}
                    \!\!  {\displaystyle{\delta H_1/\delta u}}\!\!  \\\noalign{\vskip 5pt}
                    \!\!  {\displaystyle{\delta H_1/\delta v}}\!\!  \\
                   \end{array}
                 \right)\;.
\]
As is well known from the Hamiltonian structures ${\cal D}_1$ and ${\cal D}_2$ we can construct a recursion operator $R={\cal D}_2{\cal D}_1^{-1}$ that yields the hierarchy of equations, symmetries and conserved charges associated with these two-component SP equations \cite{Olver1993}.

\section{Pietrzyk--Kanatt\v{s}ikov--Bandelow Systems}

 Pietrzyk et al. \cite{Pietrzyk2008} introduced the following three integrable two-component SP equations
\begin{eqnarray}
u_{xt} =u +{1\over 6}\left(u^3+3uv^2\right)_{xx}\;,&& v_{xt} =v +{1\over 6}\left(v^3+3u^2v\right)_{xx}\;,\label{pietrzyk1}\\
u_{xt} =u +{1\over 6}\left(u^3-3uv^2\right)_{xx}\;,&& v_{xt} =v -{1\over 6}\left(v^3-3u^2v\right)_{xx}\;,\label{pietrzyk2}\\
u_{xt} =u +{1\over 6}\left(u^3\right)_{xx}\;,&& v_{xt} =v +{1\over 2}\left(u^2v\right)_{xx}\;,\label{pietrzyk3}
\end{eqnarray}
by a direct generalization of the SP equation's ZCR matrices (\ref{zcsp}). They follow from (\ref{zc})
with $4\times 4$ matrices  given by
\begin{equation}
X=\left(
                   \begin{array}{cc}
                     \lambda\mathbb{I} & \lambda \mathbb{U}_x\\\noalign{\vskip 10pt}
                     \lambda \mathbb{U}_x & -\lambda\mathbb{I}\\
                   \end{array}
                 \right)\;,\qquad
                 T=\left(
                   \begin{array}{cc}
                     {\lambda\over2}\mathbb{U}^2+{1\over 4\lambda}\mathbb{I} & {\lambda\over6}\left(\mathbb{U}^3\right)_{x}- {1\over2}\mathbb{U}\\\noalign{\vskip 10pt}
                      {\lambda\over6}\left(\mathbb{U}^3\right)_{x}+ {1\over2}\mathbb{U}& -{\lambda\over2}\mathbb{U}^2-{1\over 4\lambda}\mathbb{I} \\
                   \end{array}
                 \right)\;,\label{zcpietrzyk}
\end{equation}
where $\mathbb{I}$ is the $2\times 2$ identity matrix and
\[
\mathbb{U}=\left(\begin{array}{cc}
                    u & v\\\noalign{\vskip 10pt}
                     v & u\\
                   \end{array}
                    \right)\;,
                    \qquad
                    \mathbb{U}=\left(\begin{array}{cc}
                    u & v\\\noalign{\vskip 10pt}
                     -v & u\\
                   \end{array}
                    \right)\;,
                    \qquad
                    \mathbb{U}=\left(\begin{array}{cc}
                    u & v\\\noalign{\vskip 10pt}
                     0 & u\\
                   \end{array}
                    \right)\;,
\]
reproduce  (\ref{pietrzyk1}), (\ref{pietrzyk2}) and (\ref{pietrzyk3}), respectively.
In \cite{Sakovich2008} Sakovich pointed out that equations (\ref{pietrzyk1}) and (\ref{pietrzyk2}) can be reduced to two decoupled SP equations (\ref{SP}) in the new variables $(p=u+v,q=u-v)$ and $(p=u+iv,q=u-iv)$, respectively. So, (\ref{pietrzyk1}) and (\ref{pietrzyk2}) are not a short pulse counterpart of the Manakov system \cite{Manakov1974} of coupled NLS equations where the polarization modes do interact nonlinearly \cite{Ablowitz2004}. For completeness we will obtain the bi-Hamiltonian structure for (\ref{pietrzyk1}) and (\ref{pietrzyk2}). These equations written simultaneously in a nonlocal form
\begin{equation}
u_t=\partial^{-1}u+{1\over6}\left(u^3\pm 3uv^2\right)_{x}\;,\qquad v_t=\partial^{-1}v\pm{1\over6}\left(v^3\pm 3u^2v\right)_{x}\;,\label{nlpietrzyk12}
\end{equation}
follow from the following Lagrangian density
\[
{\cal L}=v\left(\partial^{-1}u_t\right) - {1\over6}u^3v \mp {1\over6}v^3u + \left(\partial^{-1}u\right)\left(\partial^{-1}v\right)\;,
\]
and we can read the first  Hamiltonian structure directly,
\begin{equation}
{\cal D}_1=\left(
                   \begin{array}{cc}
                     0 & {\partial} \\
                     {\partial} & 0 \\
                   \end{array}
                 \right)\;,
\qquad H_2=\int dx\left[{1\over6}u^3v\pm{1\over6}v^3u-(\partial^{-1}u)(\partial^{-1}v)\right]\,.\label{fhspietrzyk12}
\end{equation}
From the diagonal second Hamiltonian structure for the decoupled system with elements given by (\ref{shssp}) in the variables $p$ and $q$
\[
{\cal D}_2=\left(
                   \begin{array}{cc}
                     \partial^{-1}+p_x\partial^{-1}p_x & 0 \\\noalign{\vskip 10pt}
                     0 & \partial^{-1}+q_x\partial^{-1}q_x \\
                   \end{array}
                 \right)\;,
\qquad H_1={1\over2}\int dx\left(p^2+ q^2\right)\,.
\]
we use the usual results \cite{Antonowicz1990} for  transformation laws of Hamiltonian structures under the above transformations $(p,q)\to(u,v)$ to get
\begin{equation}
{\cal D}_2=\left(
                   \begin{array}{cc}
                  \partial^{-1}+u_x\partial^{-1}u_x\pm v_x\partial^{-1}v_x&u_x\partial^{-1}v_x + v_x\partial^{-1}u_x \\\noalign{\vskip 10pt}
                    u_x\partial^{-1}v_x +v_x\partial^{-1}u_x & \pm\partial^{-1}\pm u_x\partial^{-1}u_x\pm v_x\partial^{-1}v_x\\
                   \end{array}
                 \right)\;,
\qquad H_1={1\over2}\int dx\left(u^2\pm v^2\right)\,.\label{shspietrzyk12}
\end{equation}
The proof of the Jacobi identity for these structures (and the next ones) as well their compatibility necessary to satisfy Magri's theorem \cite{Magri1978} can be proved by the standard method of prolongation described in \cite{Olver1993} in a straightforward way.

Equations (\ref{pietrzyk3}) written in a nonlocal form
\begin{equation}
u_t=\partial^{-1}u+{1\over6}\left(u^3\right)_{x}\;,\qquad v_t=\partial^{-1}v+{1\over2}\left(u^2v\right)_{x}\;,\label{nlpietrzyk3}
\end{equation}
can be derived from the following Lagrangian density
\[
{\cal L}=v\left(\partial^{-1}u_t\right) - {1\over6}u^3v +\left(\partial^{-1}u\right)\left(\partial^{-1}v\right)\;,
\]
which gives the first  Hamiltonian structure
\begin{equation}
{\cal D}_1=\left(
                   \begin{array}{cc}
                     0 & {\partial} \\
                     {\partial} & 0 \\
                   \end{array}
                 \right)\;,
\qquad H_2=\int dx\left[{1\over6}u^3v-(\partial^{-1}u)(\partial^{-1}v)\right]\,.\label{fhspietrzyk3}
\end{equation}
Also, was remarked in  \cite{Pietrzyk2008} that the system (\ref{pietrzyk3}) describes the propagation of a small perturbation $v$ on the background of a solution $u$ of the scalar SP equation (\ref{SP}) and from this perspective  we can rewrite (\ref{pietrzyk3}) as
\begin{equation}
u_{xt} =u +{1\over 6}\left(u^3\right)_{xx}\equiv K(u)\;,\quad v_{xt} =K'(u)[v]\;,\label{pietrzykuv}
\end{equation}
where
\begin{equation}
K'(u)[v]={d\ \over d\epsilon}K(u+\epsilon v)\Big|_{\epsilon=0}\label{frechet}
\end{equation}
is the Fr\'{e}chet or directional derivative of $K(u)$ in the direction of  $v$. So, the equation for $v$ is the linearized equation for $u$ given by (\ref{SP}) and objects for the system (\ref{pietrzyk3}) or (\ref{pietrzykuv}) such as conserved charges and Hamiltonian structures can be obtained directly from the corresponding objects of the scalar SP equation (\ref{SP}) using the formulas given in \cite{Kupershmidt1986} such as
 \[
 {\cal D}=\left(
                   \begin{array}{cc}
                     0 & {\widetilde{\cal D}} \\\noalign{\vskip 5pt}
                     {\widetilde{\cal D}} & {\widetilde{\cal D}}'(u)[v] \\
                   \end{array}
                 \right)\;,
                 \qquad
                 H={\widetilde H}'(u)[v]\;,
 \]
 where the ``$\sim$" referes to the objects of the scalar SP system (\ref{SP}). Hence,
 the second Hamiltonian structure for (\ref{nlpietrzyk3}) can be obtained from (\ref{shssp})  (as well the first one (\ref{fhspietrzyk3}) from (\ref{fhssp})) and it is
\begin{equation}
{\cal D}_2=\left(
                   \begin{array}{cc}
                     0 &\partial^{-1}+u_x\partial^{-1}u_x \\\noalign{\vskip 10pt}
                     \partial^{-1}+u_x\partial^{-1}u_x & v_x\partial^{-1}u_x +u_x\partial^{-1}v_x\\
                   \end{array}
                 \right)\;,
\qquad H_1=\int dx\;uv\,.\label{shspietrzyk3}
\end{equation}

\section{Dimakis--M{\"{u}}ller-Hoissen--Matsuno System}

 From a bidifferential approach to the AKNS hierarchies \cite{Dimakis2010} or from a bilinear formalism combined with a hodograph transformation \cite{Matsuno2011} these authors proposed the multi-component system
\[
u_{i,xt}=u_i+{1\over2}\left( u_{i,x}\sum_{j=1}^n u_j^2\right)_x\;,\quad i=1,2,\dots n\;,
\]
with reduced two-component SP system given by
\[
u_{xt} =u +{1\over 2}\left[(u^2+v^2)u_x\right]_{x}\;, \qquad v_{xt} =v +{1\over 2}\left[(u^2+v^2)v_x\right]_{x}\;,
\]
where $u=u_1$ and $v=u_2$. After  the transformations $(u,v)\to({(u+v)/2},{(u-v)/2i})$ this system can be written as
\begin{equation}
u_{xt} =u +{1\over 2}\left(uvu_x\right)_{x}\;, \qquad v_{xt} =v +{1\over 2}\left(uvv_x\right)_{x}\;,\label{matsuno}
\end{equation}
which is integrable with zero curvature (\ref{zc}) with matrices  given by
\begin{equation}
X=\left(
                   \begin{array}{cc}
                     \lambda & \lambda u_x\\\noalign{\vskip 10pt}
                     \lambda v_x & -\lambda\\
                   \end{array}
                 \right)\;,\qquad
                 T=\left(
                   \begin{array}{cc}
                     {\lambda\over2}uv+{1\over 4\lambda} & {\lambda\over2}uvu_x- {1\over2}u\\\noalign{\vskip 10pt}
                     {\lambda\over2}uvv_x + {1\over2}v& -{\lambda\over2}uv-{1\over 4\lambda} \\
                   \end{array}
                 \right)\;.\label{zcmatsuno}
\end{equation}
A Lagrangian density for the equations (\ref{matsuno}) written in a nonlocal form can be easily obtained and it is
\[
{\cal L}=uv_t-u(\partial^{-1}v)-{1\over 8}u^2(v^2)_x\;,
\]
and we can read the first  Hamiltonian structure as
\begin{equation}
{\cal D}_1=\left(
                   \begin{array}{cc}
                     0 & {-1} \\
                     {1} & 0 \\
                   \end{array}
                 \right)\;,
\qquad H_2=\int dx\left[u(\partial^{-1}v)+{1\over8}u^2(v^2)_x\right]\,.\label{fhsmatsuno}
\end{equation}

To find the recursion operator of the  system (\ref{matsuno}), we use the cyclic basis method \cite{Sakovich1995,Sakovich2004}. We pose the problem of finding all the evolutionary systems
\begin{equation}
u_t = f\;, \qquad v_t = g\;, \label{evol-sys}
\end{equation}
possessing the ZCR (\ref{zc}) with the matrix $X$ given by (\ref{zcmatsuno}) and the matrix $T$ being initially not fixed. We rewrite the ZCR in its equivalent (covariant, or characteristic) form
\begin{equation}
f \, C_u + g \, C_v = \nabla_x S\;,\label{char-zcr}
\end{equation}
where $C_u$ and $C_v$ are the characteristic matrices,
\begin{equation}
C_u = - \nabla_x \left( {\partial X \over \partial u_x} \right) , \qquad C_v = - \nabla_x \left( {\partial X \over \partial v_x} \right)\;,
\end{equation}
since $X = X ( u_x , v_x )$. The covariant derivative operator $\nabla_x$ is defined as $\nabla_x H = \partial H - [ X , H ]$ for any $2 \times 2$ matrix $H$, and the matrix $S$ is related to the unknown matrix $T$ as
\begin{equation}
S = T - f \, {\partial X \over \partial u_x} - g \, {\partial X \over \partial v_x}\;.
\end{equation}
For $X$ given by (\ref{zcmatsuno}) we find that the cyclic basis is three-dimensional and consists of the matrices
\begin{gather}
C_u =
\begin{pmatrix}
- \lambda^2 v_x & 2 \lambda^2 \\\noalign{\vskip 10pt}
0 & \lambda^2 v_x
\end{pmatrix}
\;, \qquad
C_v =
\begin{pmatrix}
\lambda^2 u_x & 0 \\\noalign{\vskip 10pt}
- 2 \lambda^2 & - \lambda^2 u_x
\end{pmatrix}
\;, \notag \\ \notag\\
\nabla_x C_u =
\begin{pmatrix}
2 \lambda^3 v_x - \lambda^2 v_{xx} & - 2 \lambda^3 ( 2 + u_x v_x ) \\\noalign{\vskip 10pt}
2 \lambda^3 v_x^2 & - 2 \lambda^3 v_x + \lambda^2 v_{xx}
\end{pmatrix}
\;,
\end{gather}
and the closure equations for the basis being
\begin{gather}
\nabla_x C_v = a_1 C_u + a_2 C_v + a_3 \nabla_x C_u\; , \notag \\\noalign{\vskip 8pt}
\nabla_x^2 C_u = b_1 C_u + b_2 C_v + b_3 \nabla_x C_u\; , \label{clo-eqns}
\end{gather}
where
\begin{eqnarray}
&& a_1 = \lambda\left[ u_x^2 - \frac{ ( 2 + u_x v_x ) u_{xx}}{v_{xx}}\right]\;,\qquad
a_2 = \lambda \left[( 2 + u_x v_x ) -  \frac{v_x^2 u_{xx}}{v_{xx}}\right]\;, \qquad a_3 = - \frac{u_{xx}}{v_{xx}}\;, \notag \\\noalign{\vskip 10pt}
&& b_1 =-\lambda\left[ v_x u_{xx} + 2 u_x v_{xx} - \frac{ ( 2 + u_x v_x ) v_{xxx}}{v_{xx}}\right]+ \lambda^2\left[\frac{ ( 2 + u_x v_x ) v_x^2 u_{xx}}{v_{xx}} - u_x^2 v_x^2 \right] \;, \notag \\\noalign{\vskip 10pt}
&& b_2 =-\lambda\left[3 v_x v_{xx} - \frac{ v_x^2 v_{xxx}}{v_{xx}}\right]+ \lambda^2\left[\frac{ v_x^4 u_{xx}}{v_{xx}} - ( 2 + u_x v_x ) v_x^2\right]  \;, \notag \\\noalign{\vskip 10pt}
&& b_3 = - \lambda\left[ ( 2 + u_x v_x ) -\frac{ v_x^2 u_{xx}}{v_{xx}}\right] + \frac{v_{xxx}}{v_{xx}}\; . \label{a1--b3}
\end{eqnarray}
We decompose the matrix $S$ over the cyclic basis as
\begin{equation}
S = p \, C_u + q \, C_v + r \, \nabla_x C_u\;,
\end{equation}
and we obtain from \eqref{char-zcr} and \eqref{clo-eqns} that
\begin{eqnarray}
&&f = \partial p + a_1 q + b_1 r\;, \notag \\\noalign{\vskip 5pt}
&&g = \partial q + a_2 q + b_2 r\;, \notag \\\noalign{\vskip 5pt}
&&p = - \partial r - a_3 q - b_3 r\;,
\end{eqnarray}
which can be rewritten, using \eqref{a1--b3}, as
\begin{equation}
\begin{pmatrix}
f \\\noalign{\vskip 7pt} g
\end{pmatrix}
= \left( M + \lambda L + \lambda^2 K \right)
\begin{pmatrix}
q \\\noalign{\vskip 7pt} r
\end{pmatrix}
\;, \label{op-form}
\end{equation}
where the $2 \times 2$ matrix differential operators $M$, $L$ and $K$ do not contain the spectral parameter $\lambda$, and their no null components are
\begin{eqnarray}
&&M_{11} = \frac{u_{xx}}{v_{xx}} \partial + \frac{u_{xxx}}{v_{xx}} - \frac{u_{xx} v_{xxx}}{v_{xx}^2}\;, \qquad
M_{12} = - \partial^2 - \frac{v_{xxx}}{v_{xx}} \partial + \frac{v_{xxx}^2}{v_{xx}^2} - \frac{v_{xxxx}}{v_{xx}}\;,\qquad M_{21} = \partial\;, \notag \\\noalign{\vskip 5pt}
&&L_{11} = u_x^2 - \frac{( 2 + u_x v_x ) u_{xx}}{v_{xx}}\;,\qquad L_{21} = 2 + u_x v_x - \frac{v_x^2 u_{xx}}{v_{xx}} , \qquad L_{22} = - 3 v_x v_{xx} + \frac{v_x^2 v_{xxx}}{v_{xx}}\;, \notag \\\noalign{\vskip 5pt}
&&L_{12} = \left( 2 + u_x v_x - \frac{v_x^2 u_{xx}}{v_{xx}} \right) \partial - 2 v_x u_{xx} - u_x v_{xx} -  \frac{v_x^2 u_{xxx}}{v_{xx}} + \left( 2 + u_x v_x + \frac{v_x^2 u_{xx}}{v_{xx}} \right) \frac{v_{xxx}}{v_{xx}}\;, \notag \\\noalign{\vskip 5pt}
&&K_{12} = - u_x^2 v_x^2 + \frac{( 2 + u_x v_x ) v_x^2 u_{xx}}{v_{xx}} \;, \qquad K_{22} = - ( 2 + u_x v_x ) v_x^2 + \frac{v_x^4 u_{xx}}{v_{xx}}\;.\label{comp-mlk}
\end{eqnarray}
The hierarchy of evolutionary systems \eqref{evol-sys} possessing ZCR (\ref{zc}) with the matrix $X$ given by (\ref{zcmatsuno}) is completely determined by the relation \eqref{op-form}, where $q$ and $r$ must satisfy the conditions $\partial f / \partial \lambda = \partial g / \partial \lambda = 0$. If we would have $K = 0$ in \eqref{op-form}, we could immediately write down the recursion operator for the represented hierarchy as $R = M L^{-1}$ \cite{Sakovich2004}. The more complicated case of the relation \eqref{op-form} with $K \neq 0$ was studied in \cite{Karasu2004}, where it was shown that the recursion operator $R$ is given by
\begin{equation}
R = M N^{-1}\;, \qquad N = M L^{-1} K - L\;, \label{expr-rn}
\end{equation}
if the operators $L$ and $K$ satisfy the condition
\begin{equation}
K L^{-1} K = 0\; . \label{cond-klk}
\end{equation}
In the present case of $X$  it turns out that the operators $L$ and $K$  do satisfy the condition \eqref{cond-klk}, and we have from \eqref{comp-mlk} and \eqref{expr-rn} the following:
\begin{equation}
M =
\begin{pmatrix}
\partial \dfrac{u_{xx}}{v_{xx}} & - \partial \dfrac{1}{v_{xx}} \partial v_{xx} \\[8pt]\noalign{\vskip 10pt}
\partial & 0
\end{pmatrix}\;,\quad N =
\begin{pmatrix}
( 2 + u_x v_x ) \dfrac{u_{xx}}{v_{xx}} - u_x^2 & - ( 2 + u_x v_x ) \dfrac{1}{v_{xx}} \partial v_{xx} + u_x v_{xx} \\[8pt]\noalign{\vskip 10pt}
- ( 2 + u_x v_x ) + \dfrac{v_x^2 u_{xx}}{v_{xx}} & - \dfrac{v_x^2}{v_{xx}} \partial v_{xx} + v_x v_{xx}
\end{pmatrix}
\;,
\end{equation}
and then
\[
{R}=\left(
                   \begin{array}{cc}
                    {1\over2}\partial-{1\over4}\partial Fu_x\partial^{-1}Fv_x\partial & -{1\over4}\partial Fu_x\partial^{-1}Fu_x\partial \\\noalign{\vskip 15pt}
                    {1\over4}\partial Fv_x\partial^{-1}Fv_x\partial  & -{1\over2}\partial+{1\over4}\partial Fv_x\partial^{-1}Fu_x\partial \\
                   \end{array}
                 \right)\;,
\]
where
\[
F=\left(1+u_xv_x\right)^{-1/2}\;.
\]
Its inverse can be easily obtained to yield
\[
R^{-1}=\left(
                   \begin{array}{cc}
                     2\partial^{-1}+u_x\partial^{-1}v_x & -u_x\partial^{-1}u_x \\\noalign{\vskip 15pt}
                    v_x\partial^{-1}v_x  & -2\partial^{-1}-v_x\partial^{-1}u_x \\
                   \end{array}
                 \right)\;,
\]
and from the factorization $R^{-1}={\cal D}_2{\cal D}_1^{-1}$ and the first structure (\ref{fhsmatsuno}) we get the second Hamiltonian structure of the Matsuno system
\[
{\cal D}_2=\left(
                   \begin{array}{cc}
                     u_x\partial^{-1}u_x & 2\partial^{-1}+u_x\partial^{-1}v_x \\\noalign{\vskip 10pt}
                      2\partial^{-1}+v_x\partial^{-1}u_x  & v_x\partial^{-1}v_x\\
                   \end{array}
                 \right)\;,
\qquad H_1={1\over 2}\int dx\,uv\,.
\]

\section{Feng System}

Using two sets of bilinear equations of a two-dimensional Toda lattice  linked by a B{\"a}cklund transformation
\[
\left\{ \begin{array}{l}
\!\!\!D_sD_yf\cdot f={1\over2}(f^2-\bar{f}^2)\;,\\\noalign{\vskip 10pt}
\!\!\!D_sD_y\bar{f}\cdot \bar{f}={1\over2}(\bar{f}^2-{f}^2)\;,
\end{array} \right.
\quad
\left\{ \begin{array}{l}
\!\!\!D_sD_yg\cdot g={1\over2}(g^2-\bar{g}^2)\;,\\\noalign{\vskip 10pt}
\!\!\!D_sD_y\bar{g}\cdot \bar{g}={1\over2}(\bar{g}^2-{g}^2)\;,
\end{array} \right.
\]
a particular hodograph transformation
 \[
 \left\{ \begin{array}{l}
\!\!\!x=y-\left(\ln(F\bar{F})\right)_s\;,\\\noalign{\vskip 2pt}
\!\!\!t=s\;,
\end{array} \right.
 \]
 and the dependent variable transformation
 \[
 u=i\left(\ln{\bar{F}\over F}\right)_s\;,\quad v=i\left(\ln{\bar{G}\over G}\right)_s\;,
 \]
 where $F=fg$, $G=f\bar{g}$ and $\bar{F}$ and $\bar{G}$ are the complex conjugate of $F$ and $G$ respectively,  Feng \cite{Feng2012} proposed the coupled SP equation
\begin{equation}
u_{xt} =u +uu_x^2+{1\over 2}\left(u^2+v^2\right)u_{xx}\;, \qquad v_{xt} =v +vv_x^2+{1\over 2}\left(u^2+v^2\right)v_{xx}\;.\label{feng}
\end{equation}
From a prolongation study \cite{Wahlquist1975} of this system we have obtained for the zero curvature (\ref{zc}) the matrices
\begin{eqnarray}
&&X=\left(
                   \begin{array}{cc}
                     \lambda(1+u_xv_x) & \lambda(u_x-v_x)\\\noalign{\vskip 10pt}
                     \lambda(u_x-v_x) & -\lambda(1+u_xv_x)\\
                   \end{array}
                 \right)\;,\nonumber\\\noalign{\vskip 10pt}
              &&   T=\left(
                   \begin{array}{cc}
                     {\lambda\over2}(u^2+v^2)(1+u_xv_x)+{1\over 4\lambda} & {\lambda\over2}(u^2+v^2)(u_x-v_x)- {1\over2}(u-v)\\\noalign{\vskip 10pt}
                     {\lambda\over2}(u^2+v^2)(u_x-v_x)+ {1\over2}(u-v) & -{\lambda\over2}(u^2+v^2)(1+u_xv_x)-{1\over 4\lambda} \\
                   \end{array}
                 \right)\;.\label{zcfeng}
\end{eqnarray}
Equations (\ref{feng}) written in a nonlocal form can be derived from the following Lagrangian density
\[
{\cal L}=u_{xt}v-uv-{1\over6}v^3u_{xx}-{1\over6}u^3v_{xx}\;,
\]
and we can read the following first  Hamiltonian structure
\begin{equation}
{\cal D}_1=\left(
                   \begin{array}{cc}
                     0 & \partial^{-1} \\
                     \partial^{-1} & 0 \\
                   \end{array}
                 \right)\;,
\qquad H_2=\int dx\,\left(uv+{1\over 6}u^3v_{xx}+{1\over 6}v^3u_{xx}\right)\,.\label{fhsfeng}
\end{equation}

It is not possible to obtain a recursion operator from the matrix $X$ in (\ref{zcfeng}) by cyclic basis technique, for the same reason as the Example 6 in \cite{Sakovich1995}. Namely, the transformation (\ref{sakovichsigpr})--(\ref{sakovichpq}), given below, changes the linear problem (\ref{spsp}) with the matrices (\ref{zcfeng}) into the linear problem $\Psi_y = Y \Psi$ and $\Psi_s = S \Psi$, where $\Psi(y,s) = \Phi(x,t)$, with the matrices
\begin{equation}
Y =
\begin{pmatrix}
\lambda \cos \sigma' & \lambda \sin \sigma' \\
\lambda \sin \sigma' & - \lambda \cos \sigma'
\end{pmatrix}
, \qquad S =
\begin{pmatrix}
\frac{1}{4 \lambda} & - \frac{1}{2} \sigma'_s \\[4pt]
\frac{1}{2} \sigma'_s & - \frac{1}{4 \lambda}
\end{pmatrix}
, \notag
\end{equation}
where two dependent variables $u$ and $v$ have merged into one dependent variable $\sigma'$. However, we can derive the Hamiltonian structures for the Feng system from the well-known ones of the sine-Gordon (SG) equation $\sigma_{ys} = \sin \sigma$, in light cone coordinates \cite{Case1982},
\begin{equation} \label{sakovichsgh1}
\mathcal{D}^{(\sigma)} = \partial_y^{-1} , \qquad H^{(\sigma)} = - \int \!\! dy \cos \sigma ,
\end{equation}
and
\begin{equation} \label{sakovichsgh2}
\mathcal{D}^{(\sigma)} = \partial_y + \sigma_y \partial_y^{-1} \sigma_y , \qquad H^{(\sigma)} = - \frac{1}{2} \int \!\! dy \cos \sigma \left( \partial_y^{-1} \sin \sigma \right)^2 ,
\end{equation}
using the transformation $(x,t,u,v)\to(y,s,\sigma,\sigma')$ between the Feng system and the pair of uncoupled SG equations, $\sigma_{ys} = \sin \sigma$ and $\sigma_{ys}' = \sin \sigma'$, discovered in \cite{Feng2012}. In fact, the Feng system (\ref{feng}) and the pair of SG equations in evolutionary form
\begin{equation} \label{sakovichevol}
\partial_t \!
\begin{pmatrix}
u \\ v
\end{pmatrix}
=
\begin{pmatrix}
f \\ g
\end{pmatrix}
, \qquad
\partial_s \!
\begin{pmatrix}
\sigma \\ \sigma'
\end{pmatrix}
=
\begin{pmatrix}
h \\ h'
\end{pmatrix}
,
\end{equation}
where
\begin{equation*}
\begin{split}
f & = \frac{1}{2} \left( u^2 + v^2 \right) u_x + \partial_x^{-1} ( u - v u_x v_x ) , \\
g & = \frac{1}{2} \left( u^2 + v^2 \right) v_x + \partial_x^{-1} ( v - u u_x v_x ) , \\\noalign{\vspace{5pt}}
h & = \partial_y^{-1} \sin \sigma , \qquad h' = \partial_y^{-1} \sin \sigma' ,
\end{split}
\end{equation*}
 are related to each other by the Feng's transformation
\begin{align}
\sigma & = \arccos \frac{1}{p} + \arccos \frac{1}{q} , \label{sakovichsigma} \\
\sigma' & = \arccos \frac{1}{p} - \arccos \frac{1}{q} , \label{sakovichsigpr} \\
t & = s , \qquad x = z(y,s): \; z_y = \frac{1}{p q} , \label{sakovichtx} \\\noalign{\vspace{5pt}}
p^2 & = 1 + u_x^2 , \qquad q^2 = 1 + v_x^2 . \label{sakovichpq}
\end{align}

Under the transformation \eqref{sakovichsigma}--\eqref{sakovichpq} the right-hand sides of evolution systems \eqref{sakovichevol} are related to each other as
\begin{equation} \label{sakovichfgp}
\begin{pmatrix}
f \\ g
\end{pmatrix}
= P
\begin{pmatrix}
h \\ h'
\end{pmatrix}
,
\end{equation}
with a $2 \times 2$ matrix linear operator $P$. Applying $\partial_s$ to \eqref{sakovichsigma} and \eqref{sakovichsigpr}, we get
\begin{equation} \label{sakovichhhpr}
\begin{split}
h & = \frac{1}{p^2} ( u_{xx} z_s + \partial_x f ) + \frac{1}{q^2} ( v_{xx} z_s + \partial_x g ) , \\
\noalign{\vspace{5pt}}
h' & = \frac{1}{p^2} ( u_{xx} z_s + \partial_x f ) - \frac{1}{q^2} ( v_{xx} z_s + \partial_x g ) .
\end{split}
\end{equation}
Applying $\partial_s$ to the relation $z_y = 1 / ( p q )$ of \eqref{sakovichtx} and taking into account that
\begin{equation*}
\partial_y = \frac{1}{p q} \partial_x , \qquad \partial_y^{-1} = \partial_x^{-1} p q ,
\end{equation*}
we get
\begin{equation} \label{sakovichzs}
z_s = - \frac{1}{2} \partial_x^{-1} \left[ ( u_x + v_x ) h + ( u_x - v_x ) h' \right] .
\end{equation}
Eliminating $z_s$ from \eqref{sakovichhhpr} and \eqref{sakovichzs}, we obtain \eqref{sakovichfgp} with
\begin{equation} \label{sakovichmatp}
P = \frac{1}{2} \partial_x^{-1} \!
\begin{pmatrix}
p^2 + u_{xx} \partial_x^{-1} ( u_x + v_x ) &
p^2 + u_{xx} \partial_x^{-1} ( u_x - v_x ) \\[10pt]
q^2 + v_{xx} \partial_x^{-1} ( u_x + v_x ) &
- q^2 + v_{xx} \partial_x^{-1} ( u_x - v_x )
\end{pmatrix}
.
\end{equation}

Let us consider the systems \eqref{sakovichevol} as Hamiltonian ones, that is
\begin{equation*}
U_t = \mathcal{D}^{(u,v)} \Delta^{(u,v)} H^{(u,v)} , \qquad \Sigma_s = \mathcal{D}^{(\sigma,\sigma')} \Delta^{(\sigma,\sigma')} H^{(\sigma,\sigma')} ,
\end{equation*}
where
\begin{equation*}
U =
\begin{pmatrix}
u \\ v
\end{pmatrix}
, \qquad \Sigma =
\begin{pmatrix}
\sigma \\ \sigma'
\end{pmatrix}
, \qquad \Delta^{(u,v)} =
\begin{pmatrix}
{\delta}/{\delta u} \\[4pt] {\delta}/{\delta v}
\end{pmatrix}
, \qquad \Delta^{(\sigma,\sigma')} =
\begin{pmatrix}
{\delta}/{\delta \sigma} \\[4pt] {\delta}/{\delta \sigma'}
\end{pmatrix}
,
\end{equation*}
$\mathcal{D}$ are Hamiltonian operators, $H$ are Hamiltonian functionals, and their superscripts $(u,v)$ and $(\sigma,\sigma')$ indicate which system they correspond to. We have already found that $U_t = P \Sigma_s$ with $P$ given by \eqref{sakovichmatp}. Let
\begin{equation} \label{sakovichduv}
\mathcal{D}^{(u,v)} = P \mathcal{D}^{(\sigma,\sigma')} Q ,
\end{equation}
so that
\begin{equation} \label{sakovichhuv}
\Delta^{(u,v)} H^{(u,v)} = Q^{-1} \Delta^{(\sigma,\sigma')} H^{(\sigma,\sigma')} ,
\end{equation}
with a $2 \times 2$ matrix linear operator $Q$. From the relations
\begin{align*}
0 = \partial_t H^{(u,v)} & = \int \!\! dx \, U_t^T \Delta^{(u,v)} H^{(u,v)}  = \int \!\! dy \, \frac{1}{p q} ( P \Sigma_s )^T \Delta^{(u,v)} H^{(u,v)} = \int \!\! dy \, \Sigma_s^T \frac{1}{p q} P^{\dagger} \Delta^{(u,v)} H^{(u,v)} , \\[8pt]
0 = \partial_s H^{(\sigma,\sigma')} & = \int \!\! dy \, \Sigma_s^T \Delta^{(\sigma,\sigma')} H^{(\sigma,\sigma')} = \int \!\! dy \, \Sigma_s^T Q \Delta^{(u,v)} H^{(u,v)} ,
\end{align*}
where the superscripts $T$ and $\dagger$ stand for the matrix transposition and operator conjugation, respectively, we see that we should set
\begin{equation*}
Q = \frac{1}{p q} P^{\dagger} .
\end{equation*}
Due to \eqref{sakovichmatp}, we have
\begin{equation} \label{sakovichmatq}
Q = \frac{-1}{2 p q}
\begin{pmatrix}
p^2 - ( u_x + v_x ) \partial_x^{-1} u_{xx} &
q^2 - ( u_x + v_x ) \partial_x^{-1} v_{xx} \\[10pt]
p^2 - ( u_x - v_x ) \partial_x^{-1} u_{xx} &
- q^2 - ( u_x - v_x ) \partial_x^{-1} v_{xx}
\end{pmatrix}
\partial_x^{-1} ,
\end{equation}
and the explicit expression
\begin{equation} \label{sakovichinvq}
Q^{-1} = - \partial_x
\begin{pmatrix}
\displaystyle{\frac{q}{p} \left[ 1 + u_x \partial_x^{-1} \!\! \left( \frac{u_{xx}}{p^2} + \frac{v_{xx}}{q^2} \right) \right]} &
\displaystyle{\frac{q}{p} \left[ 1 + u_x \partial_x^{-1} \!\! \left( \frac{u_{xx}}{p^2} - \frac{v_{xx}}{q^2} \right) \right]} \\[25pt]
\displaystyle{\frac{p}{q} \left[ 1 + v_x \partial_x^{-1} \!\! \left( \frac{u_{xx}}{p^2} + \frac{v_{xx}}{q^2} \right) \right]} &
\displaystyle{\frac{p}{q} \left[ - 1 + v_x \partial_x^{-1} \!\! \left( \frac{u_{xx}}{p^2} - \frac{v_{xx}}{q^2} \right) \right]}
\end{pmatrix}
\end{equation}
will also be useful.

Now, using the relations \eqref{sakovichduv} and \eqref{sakovichhuv} with \eqref{sakovichmatp}, \eqref{sakovichmatq}, and \eqref{sakovichinvq}, we can derive the Feng system's Hamiltonian structures from the ones of the scalar SG equation. We can take
\begin{equation} \label{sakovichdhss}
\mathcal{D}^{(\sigma,\sigma')} =
\begin{pmatrix}
\alpha \mathcal{D}^{(\sigma)} & 0 \\[4pt]
0 & \beta \mathcal{D}^{(\sigma')}
\end{pmatrix}
, \qquad H^{(\sigma,\sigma')} = \frac{1}{\alpha} H^{(\sigma)} + \frac{1}{\beta} H^{(\sigma')}
\end{equation}
for the pair of uncoupled SG equations, where $\alpha$ and $\beta$ are arbitrary nonzero constants, while $\mathcal{D}$ and $H$ with the superscripts $(\sigma)$ and $(\sigma')$ are given either by \eqref{sakovichsgh1} or by \eqref{sakovichsgh2}. Choosing \eqref{sakovichsgh1}, we have
\begin{equation} \label{sakovichcho1}
\mathcal{D}^{(\sigma,\sigma')} =
\begin{pmatrix}
\alpha \partial_y^{-1} & 0 \\[4pt]
0 & \beta \partial_y^{-1}
\end{pmatrix}
, \qquad H^{(\sigma,\sigma')} = - \int \!\! dy \left( \frac{1}{\alpha} \cos \sigma + \frac{1}{\beta} \cos \sigma' \right) .
\end{equation}
For the Hamiltonian functional, we obtain the following:
\begin{gather*}
\Delta^{(\sigma,\sigma')} H^{(\sigma,\sigma')} =
\begin{pmatrix}
\frac{1}{\alpha} \sin \sigma \\[10pt]
\frac{1}{\beta} \sin \sigma'
\end{pmatrix}
=
\begin{pmatrix}
\displaystyle{\frac{1}{\alpha p q} ( u_x + v_x )} \\[10pt]
\displaystyle{\frac{1}{\beta p q} ( u_x - v_x )}
\end{pmatrix}
, \\[15pt]
\Delta^{(u,v)} H^{(u,v)} = Q^{-1} \Delta^{(\sigma,\sigma')} H^{(\sigma,\sigma')} = \left( \frac{1}{\beta} - \frac{1}{\alpha} \right)
\begin{pmatrix}
v_{xx} \\ u_{xx}
\end{pmatrix}
, \\[8pt]
H^{(u,v)} = \left( \frac{1}{\alpha} - \frac{1}{\beta} \right) \int \!\! dx \, u_x v_x .
\end{gather*}
The choice of $\alpha = \beta$, which would seem most natural, is actually not allowed and without loss of generality we set
\begin{equation} \label{sakovichalbe}
\alpha = - 2 , \qquad \beta = 2
\end{equation}
to get
\begin{equation}
H_1\equiv H^{(u,v)} = - \int \!\! dx \, u_x v_x .\label{shsfengh}
\end{equation}
For the Hamiltonian operator, we obtain from \eqref{sakovichcho1} with \eqref{sakovichalbe} via \eqref{sakovichduv}, \eqref{sakovichmatp}, and \eqref{sakovichmatq} the following expression:
\begin{equation}
\mathcal{D}_2\equiv\mathcal{D}^{(u,v)} =
\begin{pmatrix}
a \, \partial_x^{-1} b - b \, \partial_x^{-1} a^{\dagger} & b \, \partial_x^{-1} d - a \, \partial_x^{-1} c^{\dagger} \\[15pt]
d \, \partial_x^{-1} b - c \, \partial_x^{-1} a^{\dagger} & c \, \partial_x^{-1} d - d \, \partial_x^{-1} c^{\dagger}\label{shsfengd}
\end{pmatrix}
,
\end{equation}
where
\begin{equation*}
\begin{aligned}
a & = \partial_x^{-1} u_{xx} \partial_x^{-1} v_x , & b & = \partial_x^{-1} + u_x \partial_x^{-1} u_x , \\[15pt]
c & = \partial_x^{-1} v_{xx} \partial_x^{-1} u_x , & d & = \partial_x^{-1} + v_x \partial_x^{-1} v_x .
\end{aligned}
\end{equation*}
Using
\[
 (bv_{xx})=(v_x+c^{\dagger}u_{xx})\;,\quad (du_{xx})=(u_x+a^{\dagger}v_{xx})\;,
\]
it is straightforward to check that (\ref{feng}) in evolutionary form follows from (\ref{shsfengh}) and (\ref{shsfengd}) as a Hamiltonian system.
In completely the same way, using the SG equation's Hamiltonian structure \eqref{sakovichsgh2} and the choice \eqref{sakovichalbe} in \eqref{sakovichdhss}, we obtain (\ref{fhsfeng}).

\section{Yao--Zeng System}

 Yao and Zeng  \cite{YaoZeng2011} claim to have constructed a new spectral problem (\ref{spsp}), with matrices $X$ and $T$  spanned by a specific loop algebra, which generates a new hierarchy of coupled SP equations having
\begin{equation}
u_{xt} =u +{1\over 6}\left(u^3\right)_{xx}\;, \qquad v_{xt} =v +{1\over 2}\left(u^2v_x\right)_{x}\;,\label{yaozeng}
\end{equation}
as its first member. However, the ZCR obtained in their paper does not reproduce (\ref{yaozeng}) as can be easily checked. Also, by a direct analysis of their ZCR, one can prove that the system (\ref{yaozeng}) cannot be associated with their spectral problem. Here we establish the integrability of (\ref{yaozeng}) in the following way. By the transformation $v\to v_x$ the Pietrzyk et al system  (\ref{pietrzyk3}) can be mapped into (\ref{yaozeng}) and particularly the ZCR matrices (\ref{zcpietrzyk}) yields, after a gauge transformation \cite{Brunelli2012},
\begin{eqnarray*}
&&X=\left(
                   \begin{array}{cccc}
                     \lambda & \lambda u_x & 0 & 0\\\noalign{\vskip 5pt}
                     \lambda u_x & -\lambda &0& 0\\\noalign{\vskip 5pt}
                     0 & \lambda v_{x} &\lambda&\lambda u_x\\\noalign{\vskip 5pt}
                     -\lambda v_{x} &0&\lambda u_x&-\lambda\\
                   \end{array}
                 \right)\;,\nonumber\\\nonumber\\\nonumber\\
              &&   T=\left(
                   \begin{array}{cccc}
                     {1\over4\lambda}+{\lambda\over2}u^2 & {\lambda\over2}u^2u_x-{1\over2}u & 0& 0 \\\noalign{\vskip 10pt}
                           {\lambda\over2}u^2u_x+{1\over2}u& -{1\over4\lambda}-{\lambda\over2}u^2 & 0 & 0\\\noalign{\vskip 10pt}
                             0&  {\lambda\over2}u^2v_x-{1\over2}v &  {1\over4\lambda}+{\lambda\over2}u^2  & {\lambda\over2}u^2u_x-{1\over2}u\\\noalign{\vskip 10pt}
                         -{\lambda\over2}u^2v_x-{1\over2}v & 0 &  {\lambda\over2}u^2u_x+{1\over2}u &  -{1\over4\lambda}-{\lambda\over2}u^2
                   \end{array}\label{zcpietrzyk3}
                 \right)\;.
\end{eqnarray*}
This corrects the ZCR and in terms of the Lie algebra spanned by the generators $\{e_1,e_2,e_3,e_4,e_5,e_6\}$ considered in \cite{YaoZeng2011} we have the following decomposition
\begin{eqnarray*}
&&X=\lambda e_1+\lambda u_x(e_2+e_3)+\lambda v_x(e_5-e_6)\;,\\\noalign{\vskip 5pt}
&&T=T_{11}e_1+T_{12}e_2+T_{21}e_3+T_{32}e_5+T_{41}e_6\;,
\end{eqnarray*}
in disagreement with the one provided by Yao and Zeng in their isospectral problem.

For completeness, the bi-Hamiltonian structure of the Yao--Zeng system (\ref{yaozeng}) in evolutionary form can be obtained from (\ref{fhspietrzyk3}) and (\ref{shspietrzyk3}) using the transformation laws of Hamiltonian structures \cite{Antonowicz1990}  under the map $(u,v)\to(u,v_x)$, and they are
\[
{\cal D}_1=\left(
                   \begin{array}{cc}
                     0 & -1 \\
                     1 & 0 \\
                   \end{array}
                 \right)\;,
\qquad H_2=\int dx\left[{1\over6}u^3v_x-(\partial^{-1}u)v\right]\,,\label{fhsyaozeng}
\]
\[
{\cal D}_2=\left(
                   \begin{array}{cc}
                     0 &-\partial^{-2}-u_x\partial^{-1}u_x\partial^{-1} \\\noalign{\vskip 10pt}
                     \partial^{-2}+\partial^{-1}u_x\partial^{-1}u_x & -\partial^{-1}\left(v_{xx}\partial^{-1}u_x -u_x\partial^{-1}v_{xx}\right)\partial^{-1}\\
                   \end{array}
                 \right)\;,
\qquad H_1=\int dx\;uv_x\,.\label{shsyaozeng}
\]

In \cite{Brunelli2012} we provide details of how the Yao--Zeng system (\ref{yaozeng}) is related with a perturbation of the SP equation (\ref{SP}).

\section{Conclusions}

In this paper, we have obtained the bi-Hamiltonian structure for some of the two-component
 SP equations proposed in the literature to take into account the propagation of polarized pulses in anisotropic media. Namely, we studied systematically  the Pietrzyk--Kanatt\v{s}ikov--Bandelow \cite{Pietrzyk2008}, Dimakis--M{\"{u}}ller-Hoissen--Matsuno \cite{Dimakis2010,Matsuno2011},  Feng \cite{Feng2012} and the Yao and Zeng  \cite{YaoZeng2011} systems. Also, we have obtained the zero-curvature representation for the Feng and Yao-Zeng equations.

 Through a Painlev{\' e} analysis of two one-parameter equations belonging to the general two-component short pulse equations for cubically nonlinear anisotropic optical fibers, introduced in \cite{Pietrzyk2008},
\[
U_{n,xt}=c_{ni}U_i+c_{nijk}(U_iU_jU_k)_{xx}\;,\quad n,i,j,k=1,2\;,
\]
where $c_{ni}$ and $c_{nijk}$ are constant coefficients determined by optical properties of the medium,
two more integrable systems besides (\ref{pietrzyk1})--(\ref{pietrzyk3})  were found in \cite{Sakovich2008}
\begin{eqnarray}
u_{xt} =u +{1\over 6}\left(u^3+uv^2\right)_{xx}\;,&& v_{xt} =v +{1\over 6}\left(v^3+u^2v\right)_{xx}\;,\label{sakovich1}\\
u_{xt} =u +{1\over 6}\left(u^3\right)_{xx}\;,&& v_{xt} =v +{1\over 6}\left(u^2v\right)_{xx}\;,\label{sakovich2}
\end{eqnarray}
where $u$ and $v$ denote the polarization components $U_1$ and $U_2$. No ZCR and bi-Hamiltonian structures are known at this point and these systems are still under investigation.

\end{document}